\documentclass[%
reprint,
 amsmath,amssymb, aps,
 prl,
]{revtex4-2}

\usepackage[T1]{fontenc}
\usepackage{inputenc}
\DeclareUnicodeCharacter{2009}{\,}
\usepackage[english]{babel}
\makeatletter
\let\l@en\l@english
\makeatother
\usepackage{graphicx}% Include figure files
\usepackage{xcolor}
\usepackage{dcolumn}% Align table columns on decimal point
\usepackage{bm}% bold math
\usepackage{physics}
\usepackage{bbm}
\usepackage{amsmath}
 \usepackage{mathrsfs}
\usepackage[mathlines]{lineno}% Enable numbering of text and display math
\usepackage{hyperref}
\hypersetup{
	colorlinks=true,
	linkcolor=blue,
	filecolor=magenta,
	urlcolor=blue,
	citecolor=blue,
}

\begin{document}

\preprint{APS/123-QED}

\title{
Localized orbitals and tunnel couplings from general confinement potentials in gate-defined quantum-dot arrays
%Orbital eigenbasis localization and tunnel coupling evaluation in arbitrary gate-defined quantum dot arrays
	% \textit{A priori} evaluation of inter-orbital tunnel couplings in arbitrary gate-defined quantum dot arrays via orbital localization
	%A natural localization of single-electron orbitals in  arbitrary physical quantum dot potentials
	} 

\author{Bohdan Khromets}
\affiliation{Institute for Quantum Computing and Department of Physics, University of Waterloo, 200 University Avenue West, Waterloo, Ontario  N2L 3G1, Canada}
\email{bohdan.khromets@uwaterloo.ca}

\author{William Chow}% 
\affiliation{Institute for Quantum Computing and Department of Physics, University of Waterloo, 200 University Avenue West, Waterloo, Ontario  N2L 3G1, Canada}
\email{w5chow@uwaterloo.ca}

\author{Jonathan Baugh}% 
\affiliation{Institute for Quantum Computing and Department of Chemistry, University of Waterloo, 200 University Avenue West, Waterloo, Ontario  N2L 3G1, Canada}
\email{baugh@uwaterloo.ca}

\date{\today}% 

\begin{abstract}
	Efficient simulation of  dense gate-defined multi-quantum-dot arrays requires accurate and scalable modeling methods, compatible with asymmetries and imperfections of realistic voltage-controlled confinement potentials.
	We present a numerical localization procedure that rotates the eigenbasis of a general one-particle effective orbital Hamiltonian into $s$-, $p$-, $d$-, $\ldots$-shells of localized orbital wavefunctions associated with individual quantum dots. The pairwise tunnel couplings between such states are computed directly as matrix elements of the Hamiltonian.
	We demonstrate this procedure on a 2D triangular Si-MOS triple-quantum-dot array by obtaining the voltage dependencies of the tunnel couplings and their distributions in the presence of disorder.
	We discuss the implications of this evaluation method on the many-body calculations, and relate it to the  experimental tunnel coupling measurements.
%	 This localized-basis approach enables scalable multi-orbital Hubbard and Configuration Interaction simulations of voltage-dependent many-electron spectra.

\end{abstract}

\maketitle

% The original introduction notes are retained below for reference but excluded
% from the typeset manuscript.

\textit{Introduction}---Quantum information processing with arrays of voltage-defined quantum dots (QDs) relies on leveraging the complex interplay of the orbital, spin, and valley degrees of freedom of electronic states in semiconductors.
%
%QD-based quantum-computer architectures include charge qubits, Loss-DiVincenzo qubits, singlet-triplet qubits, and exchange-only qubits for electrons or holes~\cite{Burkard2023Semiconductorspinqubits}.
%
% What all the major QD-based quantum computer architectures (cf.~\cite{Burkard2023Semiconductorspinqubits}) share is	highly-tunable, voltage-controlled tunnel couplings between QDs. 
%
Rapid progress from isolated devices
to multi-QD registers with high initialization, control, and readout fidelities~\cite{Volk2019Loadingquantumdot, Philips2022Universalcontrolsix, Hsiao2020EfficientOrthogonalControl} 
requires refined and sophisticated models to accurately predict their quantum-mechanical electronic behavior over large voltage operating ranges and in the presence of disorder. This is crucial, for instance, for the continuous optimal control of exchange-mediated quantum-logic operations in spin qubits~\cite{Madzik2025Operatingtwoexchange}, or for high-fidelity shuttling protocols~\cite{DeSmet2025Highfidelitysingle, Brandes2002Adiabatictransferelectrons, Kandel2021Adiabaticquantumstate}. 
Calculating tunnel coupling, the energy scale that quantifies the hybridization of orbital states in neighboring QDs, is a key ingredient in this pursuit, but is challenging to accomplish for realistic, non-idealized confinement potentials. 
%
%Tunnel coupling is a key energy scale that quantifies the hybridization of orbital states in neighboring QDs. It controls the exchange interaction used to implement native spin-qubit logic gates, influences the speed and fidelity of shuttling, and is the central parameter in extending the electrostatic capacitance model into the quantum regime~\cite{Wang2011QuantumTheoryCSD}.
%
The minimal analytical description of a double-QD with a two-level-system model  (or a four-level model for the valley states in silicon~\cite{Zhao2022MeasurementTunnelCoupling, Borjans2021Intervalley_Tunnel_Coupling}) has the same limitation as the experimental measurement techniques in the single-electron regime such as 
probability crossover (``excess charge'')~\cite{Hatano2005SingleElectronDelocalization,PioroLadriere2005ChargeSensingArtificialH2+molecule,DiCarlo2004DifferentialChargeSensing,Diepen2018Automatedtuninginter}, photon-assisted tunneling~\cite{Oosterkamp_1998, Diepen2018Automatedtuninginter}, or direct transport~\cite{Huettel2005Directcontroltunnel}. Namely, a value of tunnel coupling can be extracted only within a narrow bias window, with the detuning comparable to the tunnel coupling magnitude itself (often, $\sim \mu$eV or below).
The WKB method~\cite{Bhattacharya1982DoubleMinimumWells, Huang2018Spindecoherencetwo, 
 Platt2008WKBAnalysisTunnel} unravels the  exponential dependence of the tunnel coupling on the barrier height and dot pitch, and a weak dependence on the symmetric detuning. Being one-dimensional by construction, however, this analytical framework can only be used qualitatively to describe general two- or three-dimensional QD arrays with non-unique tunneling paths.
 The work~\cite{Hsiao2020EfficientOrthogonalControl} combines these extrapolated  WKB trends and accurate pairwise measurements to demonstrate the orthogonal control of virtualized barrier gates in a four-QD array.
 %, thus being a major step forward in the pairwise control of tunnel couplings in large multidot systems.
%
However, strong simultaneous couplings among three or more QDs require much more sophisticated approaches to the measurement and modeling of tunnel rates~\cite{Braakman2013LongdistanceCoherentCoupling}.
Higher-energy orbitals and electron–electron Coulomb interactions can substantially influence tunnel couplings, with important consequences for the spin-charge states of QD ensembles. Experimental measurements of these effects, through charge-state probability crossover profiles and the curvature of transition boundaries in charge stability diagrams~\cite{Hatano2005SingleElectronDelocalization,Diepen2018Automatedtuninginter,Zwolak2018QFlowlitedataset}, have likewise been confined to narrow voltage windows near triple points. The full configuration interaction (FCI) method with natural-orbital decomposition introduced in Ref.~\cite{Foulk2024TheoryCSDs_many_orbital_CI} provides a numerical treatment of many-electron effects on tunnel coupling, but remains applicable only to weakly coupled, nearly unbiased pairs of symmetric QDs.
The approach compatible with arbitrary \textit{disordered} electric potential landscapes---the nonequilibrium  Green's
function method from Ref.~\cite{Klos2018Calculationtunnelcouplings}---is designed for calculating a dot-to-reservoir coupling, specifically, rather than pairwise dot-to-dot couplings.
The calculation of exchange interaction, an energy scale inherently related to tunnel coupling in many-electron systems,  could lead to erroneous results  
 within the minimal Heitler-London and Hund-Mulliken approximations \cite{Chan2018validitymicroscopiccalculations} even for the idealized biquadratic  \cite{Yang2011GenericHubbardmodel}, double-Gaussian \cite{DasSarma2011Hubbardmodeldescription} or quartic
 \cite{Burkard_1999} double-QD potential, since the trial Fock-Darwin  wavefunctions do  not  account for the shape and height of the tunneling barrier.
 Exchange calculation through exact diagonalization is suitable in principle for general potentials when involving large  
 analytical  \cite{Pedersen_2007,Puerto_Gimenez_2007,Buonacorsi2020Optimizinglateralquantum,Nielsen2010Implicationssimultaneousrequirements} or single-electron Hamiltonian eigenfunction bases  \cite{Anderson2022Highprecisionreal,Jnane2025Abinitiomodeling,Foulk2024TheoryCSDs_many_orbital_CI}, but is computationally demanding, with obscure and potential-specific convergence patterns, and thus limited scalability to multi-QD systems.

In this Letter,  we establish a general mapping of a multi-QD single-electron Hamiltonian eigenspace to the localized  
$s$-, $p$-, $d$-, $\ldots$-like orbital states, with the inter-orbital tunnel couplings evaluated intuitively as the Hamiltonian matrix elements.
The construction accommodates disordered and strongly biased confinement-potential landscapes defining arbitrary numbers of QDs, and applies to  a wide selection of (quasi)$-1-$, $2-$, or $3-$dimensional material systems.
We demonstrate the method, in its generality, on a planar triangular array of inequivalent Si-MOS quantum dots and use it to resolve the voltage dependence and disorder-induced distribution of tunnel couplings in such a system. 
We discuss how the method bridges the electrostatic device simulations to the expected experimental outcomes of tunnel coupling measurements, and identify its advantages for many-body calculations within multi-orbital Hubbard and FCI frameworks.

\begin{figure*}[t]
	\centering
	\includegraphics[width=0.98\linewidth]{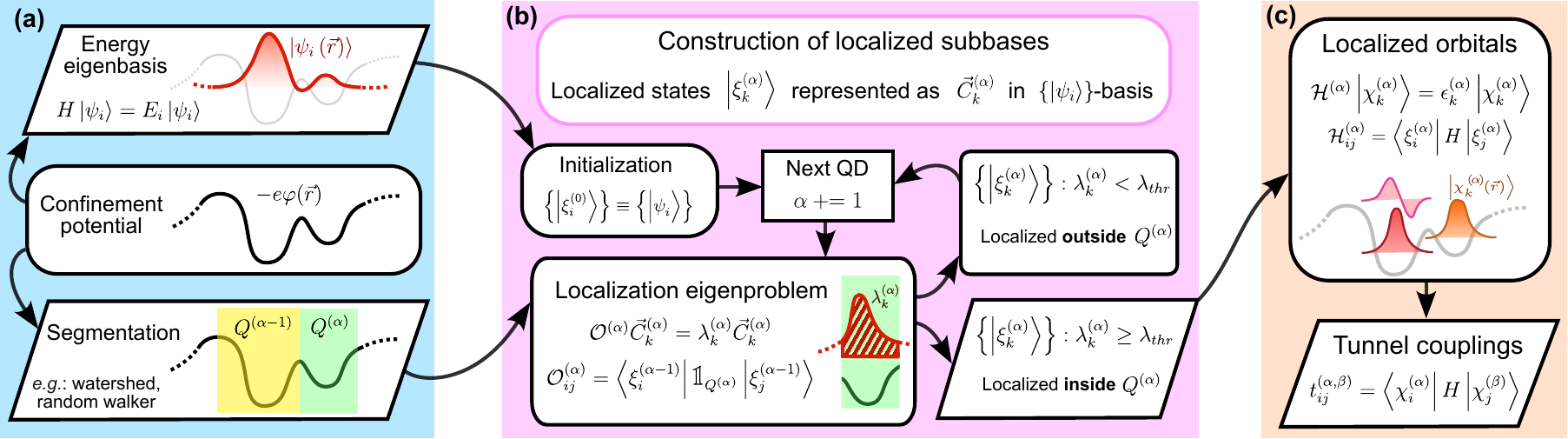}
	\caption{Localization algorithm (see End Matter for derivation). (a) Input parameters, 
	determined from the electrostatic confinement potential energy landscape $-e\varphi(\vec{r})$: 
	Hamiltonian eigenbasis $\left\{ \ket{\psi_i}\right\}$ (``$\psi$-basis''), and spatial regions
	$Q^{(\alpha)}$ of distinct QDs obtained from   a numerical segmentation algorithm. 
	(b) Decomposition of the vector space spanned by  $\left\{ \ket{\psi_i}\right\}$
	into orthogonal subspaces of states $ \left\{ \ket{\xi_i^{(\alpha)}}\right\} $, 
	 localized in distinct $Q^{(\alpha)}$ regions. 
	 % Vectors $\vec{C}_{k}^{(\alpha)}$ denote the states $\left\{ \ket{\xi_i^{(\alpha)}}\right\}$ in the $\psi$-representation and are found sequentially for different QDs one by one. 
	%
	 $\vec{C}_{k}^{(\alpha)}$ are the eigenvectors of
	 the overlap matrix $\mathcal{O}^{(\alpha )}$ of the delocalized subbasis, which coincides with the full $\psi$-basis at first iteration.
	The eigenvalues $\lambda_k^{(\alpha)}$ represent the integrated probabilities of states $\ket{\xi_k^{(\alpha)}}$  within $Q^{(\alpha)}$.  
	 The states with the highest eigenvalues, $\lambda_k^{(\alpha)}\geq \lambda_{\mathrm{thr}}$, are retained for the $\alpha^\text{th}$ QD region, whereas the subspace of states with $\lambda_k^{(\alpha)}< \lambda_{\mathrm{thr}}$ forms the delocalized basis to be decomposed over other QD regions. The procedure is run sequentially on all $Q^{(\alpha)}$.
	 (c) The  localized $s-, p-, d-, \ldots$---like orbitals $ \left\{ \ket{\chi_k^{(\alpha)}}\right\} $ of the $\alpha^\text{th}$ QD 
		 and their level energies $\epsilon_{k}^{(\alpha)}$ 
	 are found by diagonalizing the Hamiltonian projected onto the $\alpha^\text{th}$ localized subspace
	 $ \left\{ \ket{\xi_i^{(\alpha)} }\right\} $. 
	 Tunnel couplings between different localized orbitals 
	 are the matrix elements of the full system Hamiltonian in the $\chi$-basis.
	}
	\label{fig:flowchart}
\end{figure*}
\textit{Localization algorithm}---
The low-energy behavior of a single electron in a double-QD system is often described by a two-level-system (2LS) model Hamiltonian 
$H = \frac{\epsilon}{2}\tau_z + t\tau_x$ acting on the basis of two orbitals $\ket{\chi_{L/R}}$ localized in the left and right QDs, respectively. Here, $\epsilon$ is the detuning between the two QDs and $t$ is the tunnel coupling, and $\tau_z$ and $\tau_x$ are the Pauli operators in this basis.
This model accurately captures the coherent delocalization of the two lowest-energy orbitals over the two QDs only within a narrow detuning window $\abs{\epsilon}\lesssim \abs{t}$: elsewhere, the ground state collapses  towards one of the potential energy wells  (a ``flea-on-an-elephant'' effect~\cite{Simon1985FleaElephant}).  Furthermore, detunings comparable to the orbital energy spacing induce mixings with higher-energy orbitals, and the two-level system model loses its validity. 
Since any realistic  confinement potential is devoid of perfect symmetry and contains disorder, the evaluation of $t$ for an arbitrary voltage configuration by mapping the eigenenergies onto the 2LS model becomes possible only in exceptional cases of highest symmetry. %Indeed, for typical qubit processor voltage operating ranges of $10s$ of $\text{meV}$, the values of $t$ do not exceed $10s-100s$ of $\mu\text{eV}$ and are exponentially suppressed within the majority of the operating range. 
For  arrays of $N$  QDs, the minimal $N$-level-system model  not only makes small-detuning configurations rarer but also precludes any simple relationship between the system's $N$ lowest-lying eigenstates and the $N(N-1)/2$ pairwise couplings between QDs.
Note, however, that the vector space spanned by a sufficiently large number $M>N$ of lowest-lying Hamiltonian eigenstates should still contain enough information about the states localized within different QDs such as $\ket{\chi_{L/R}}$ states of the 2LS model.
If one views $\ket{\chi }$---like states as ``localized'' because they maximize the probability of being found in a specific spatial region, this extremum condition can be formalized for the full  basis of $M$  eigenstates to extract the subbases localized within different QD spatial regions. The system Hamiltonian can then be diagonalized on the corresponding vector subspaces to obtain the localized $\ket{\chi }$---like orbitals, their level energies, and their pairwise tunnel couplings. 

The algorithm for the localized-orbital-basis
construction is illustrated schematically in Fig.~\ref{fig:flowchart}.
Electrostatic confinement potential $\varphi(\vec{r})$ in $D$ spatial dimensions determines two ingredients for the localization procedure (Fig.~\ref{fig:flowchart}(a)). The first one is the basis of $M$ lowest-energy eigenstates
$\left\{ \ket{\psi_i}\right\}$ (the ``$\psi$---basis'') of the orbital Hamiltonian $H$ (the appropriate choice of $M$  will be discussed later). The second one is a set of
	distinct non-overlapping spatial regions   $\left\{ 
	Q^{(\alpha)}\right\}$  with  corresponding indicator functions $\left\{ \mathbbm{1}_{Q^{(\alpha)}}(\vec{r}) \right\}$, assigned to each QD in the array $\left( \alpha = 1\ldots N\right)$.
For the electron or hole QDs in group-IV or III-V semiconductors, 
$H$ is the effective mass Hamiltonian, and $\ket{\psi_i}$ are 
Bloch function envelope states. In the hexagonal lattice 2D materials such as graphene, or their multilayer stacks, 
$\ket{\psi_i}$ are the bound states of massless Dirac fermions, 
found from the coupled Dirac equations 
\cite{ Beenakker2008AndreevreflectionGraphene, Recher2009Boundstatesmagnetic}.  
%
	%For the simplest device geometries, it can be done heuristically by setting the regions according to the gate electrode layout and/or material interfaces. 
	For a general multi-QD voltage-dependent and/or disordered  $D$-dimensional potential landscape, QD regions $\left\{ 
	Q^{(\alpha)}\right\}$ should be extracted numerically using a $D$-dimensional data segmentation algorithm, e.g. watershed \cite{Neubert2014CompactWatershed} or random walker \cite{Grady2006_random_walker}. 
	% Watershed [CITE] and random walker [CITE] are two examples of segmentation algorithms suitable for this task: watershed floods the potential-energy landscape from its local minima and identifies watershed lines, whereas random walker treats the potential landscape as the diffusivity profile and determines the QD region to which each point in space is most likely to diffuse.
	%
	Fig. \ref{fig:flowchart}(b) illustrates the decomposition of the vector space spanned by $\left\{ \ket{\psi_i}\right\}$ into orthogonal subspaces of localized states $ \left\{ \ket{\xi_i^{(\alpha)}}\right\} $, iterated over all $Q^{(\alpha)}$ regions. As derived in the End Matter, $ \left\{ \ket{\xi_i^{(\alpha)}}\right\} $ can be found by diagonalizing the overlap matrix of a delocalized basis (full $\psi$---basis for the first iteration) and sorting the eigenvectors according to their  eigenvalues into two subbases: those localized within and outside the respective $Q^{(\alpha)}$ region. The latter subbasis is used as input in the next iteration of decomposition $\alpha\rightarrow\alpha+1$. Finally, $H$ is projected on each of the subbases localized {within} individual QDs   and diagonalized. This yields families of the localized orbital states $ \left\{ \ket{\chi_k^{(\alpha)}}\right\} $, or  ``$\chi$-states'', with level energies $\epsilon_{k}^{(\alpha)}$ (Fig.~\ref{fig:flowchart}(c)). These
	$\chi$-states form  $s-, p-, d-, \ldots$ energy shells of each  $\alpha^\mathrm{th}$ QD \cite{Leon2020spdf_electrons, Jaworowski2017MacroscopicSingletTriplet, Cockins2010EnergylevelsQDs,Vachon_2009_QD_spectrum}.
	The matrix elements of the full system Hamiltonian between distinct $\chi$-states in different QD regions then have a particularly intuitive interpretation of tunnel coupling parameters. 

\textit{Hamiltonian eigenbasis size and measurements}---A  number $M$ of Hamiltonian eigenstates $\left\{\ket{\psi_i}\right\}$ sufficient to describe at least $K$ full shells of $\chi$-orbitals per QD can be estimated  from sequentially finding probabilities $\int_{{Q}^{(\alpha)}} \abs{\psi_i(\vec{r})}^2 \dd{\vec{r}} $  for lowest-lying $\psi$-states, and retaining  $M \gtrsim M_{\min{}}$ of them,  for which the following is satisfied (see proof in the End Matter): 
	\begin{equation}\forall \alpha=1 \ldots N: \ 
 \sum_{i=1}^{M_{\min{}}} \int_{{Q}^{(\alpha)}} \abs{\psi_i(\vec{r})}^2 \dd{\vec{r}}  \gtrsim 
		\begin{pmatrix}
			\scriptstyle K-1 + D \\ \scriptstyle  D
		\end{pmatrix}
	\label{eq:minimum_subbasis_main}
\end{equation}
where the right-hand-side is a binomial coefficient.

The low-temperature tunnel coupling measurements in the single-electron regime such as probability crossover~\cite{Hatano2005SingleElectronDelocalization,PioroLadriere2005ChargeSensingArtificialH2+molecule,DiCarlo2004DifferentialChargeSensing,Diepen2018Automatedtuninginter} or photon-assisted tunneling~\cite{Oosterkamp_1998, Diepen2018Automatedtuninginter} directly probe the lowest-energy subspace near the symmetric detuning point. The minimal vector space encapsulating %tunnel couplings between 
the lowest-lying localized orbitals $\ket{\chi_0^{\alpha}}$ is captured with the choice of $K=1$ and thus should be used to estimate the experimentally  measured values: $t^{(\alpha, \beta)}\equiv t^{(\alpha, \beta)}_{00}$. 
With the growth of the  $\psi$-basis size $M$ needed for higher $K$ values, $\ket{\chi_0^{\alpha}}$ state probabilities will be progressively more concentrated within the $Q^{(\alpha)}$ region and exhibit a more rapid decay near the QD boundaries. As a result, $\abs{t^{(\alpha, \beta)}_{00}}$ parameters will gradually decrease from their values at $K=1$. Nevertheless, the higher-$K$ values give a fuller description of the Hamiltonian eigenspace and are suitable for many-electron calculations, as discussed later.  

\begin{figure*}[t]
	\centering
	\includegraphics[width=\linewidth]{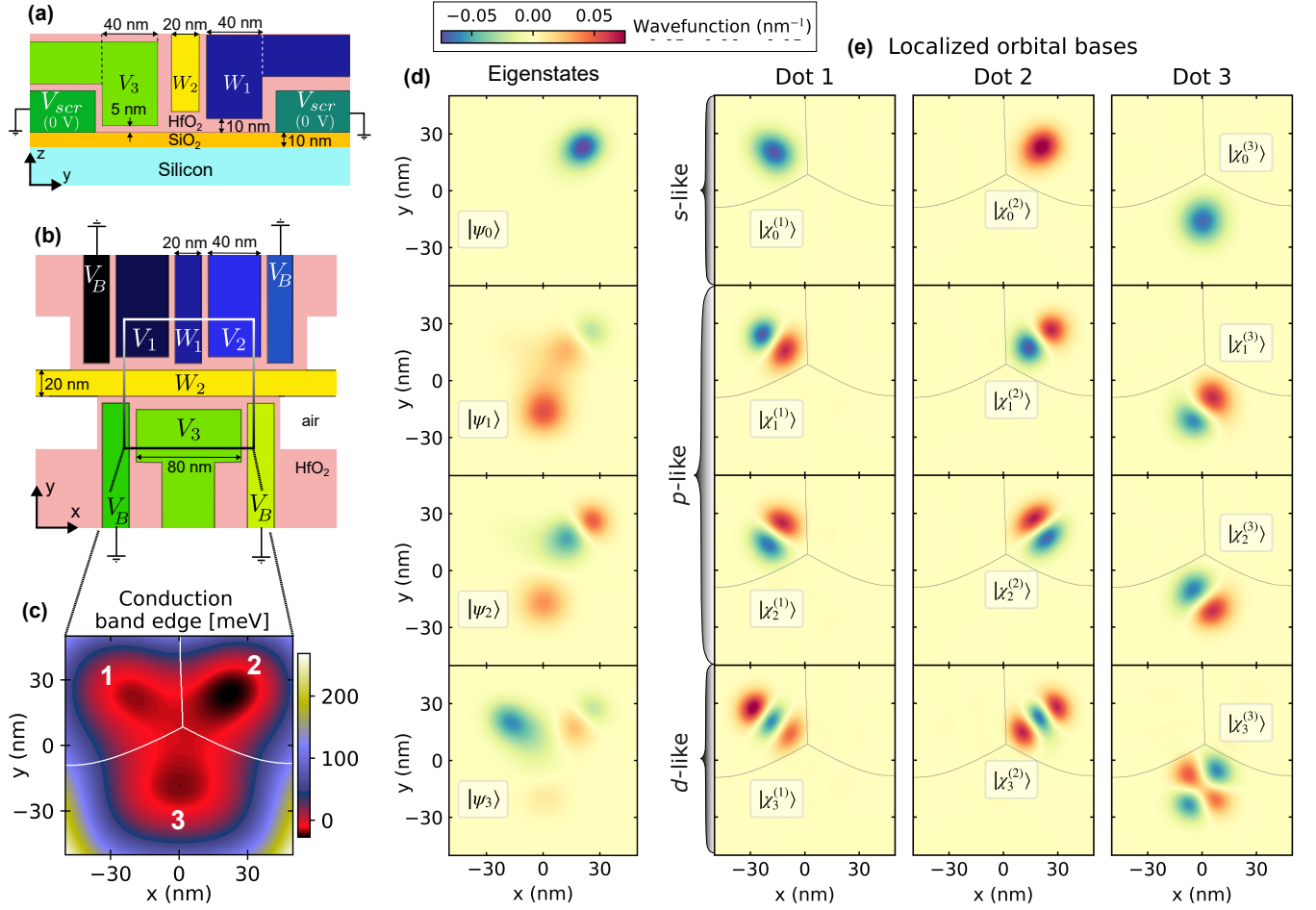}
	\caption{(a), (b) Side and top views of the triple-QD device geometry, respectively. For the data presented in this figure,
	plunger gate voltages are set to $V_1=0.98$ V, $V_2=1.02$ V, $V_3=0.925$ V, tunnel-gate voltages are set to $W_{1}=W_{2}=0.75$ V, and the barrier and screening gates are grounded, $V_B=V_{{scr}}=0$ V.
	(c) Conduction-band-edge profile $E_c(x,y)$, at
	the plane of electron accumulation.
	 White lines and labels indicate the three $Q^{(\alpha)}$ regions, obtained from 
	 the random walker segmentation of $E_c(x,y)$.
	(d) Four lowest-energy eigenstates $\ket{\psi_i(x,y)}$ of the effective-mass Hamiltonian $H$ in the conduction band landscape from (c). (e) The $s$-like, all $p$-like, and lowest-lying $d$-like orbitals $\ket{\chi_i^{(\alpha)}(x,y)}$ obtained from the localization procedure described in Fig.~\ref{fig:flowchart}.
	}
	\label{fig:gate_geometry_basis_potential}
\end{figure*}

\begin{figure*}[t]
	\centering
	\includegraphics[width=\linewidth]{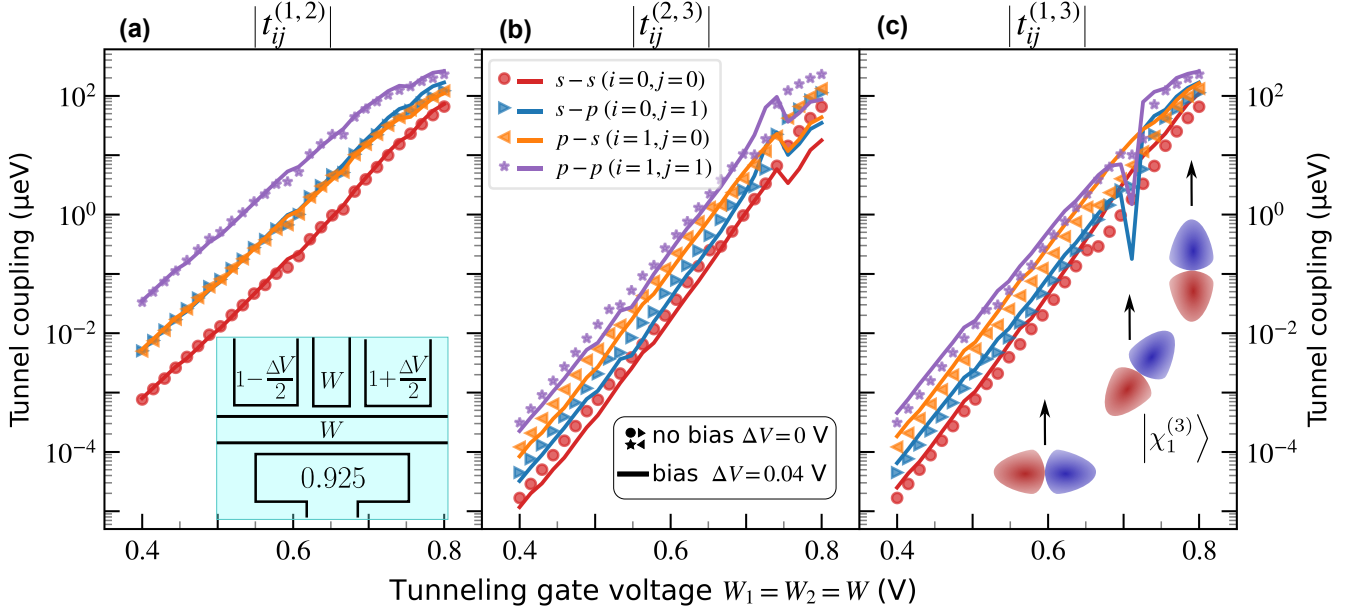}
	\caption{(a)--(c) Tunnel-coupling parameters $\abs{t^{(\alpha, \beta)}_{ij}}$ between the $s$-like and lowest-lying $p$-like orbitals ($i,j\in\{0,1\}$) for each pair of QDs in a triangular array as functions of the tunnel-gate voltage $W_1=W_2=W$. Markers indicate no bias between the first and second QDs, whereas the solid lines indicate a $40$-mV bias between them.
	The inset in (a) displays the voltage values (in volts) at the plunger and tunnel gates during the sweeps; cf. Fig.~\ref{fig:gate_geometry_basis_potential}(b). The inset in (c) depicts the orientation change of the $p$-like orbital $\ket{\chi_{1}^{(3)}}$ with voltage, which explains the nonmonotonic trends in (b,c) near $W\sim 0.7-0.73$ V.
	}
	\label{fig:tunnel_voltage_sweep}
\end{figure*}

\textit{Triple quantum dot example}---
To demonstrate our localization procedure,
we simulate a realistic multi-QD device with a high degree of connectivity, representative of the experimentally studied triangular and rectangular QD arrays in GaAs \cite{Noiri2017triangular3QD, Mukhopadhyay2018_2x2} and Si \cite{Wang2024_Si_2x2}, as well as a linear triple-QD array in the long-range tunnel-coupling regime \cite{Braakman2013LongdistanceCoherentCoupling}.
A systematic study of distinct voltage operating regimes in a low-symmetry and, possibly, disordered device that is not amenable to a two-level-system mapping or an unambiguous application of the WKB approximation will reveal both the tunnel-coupling behavior beyond the narrow measurement windows accessible to experimental techniques
and the fundamental structure of the confinement-potential-specific localized orbitals underlying these tunnel couplings.
Specifically,
we consider a  triangular array of three QDs in a gated Si-SiO${}_2$-HfO${}_2$ heterostructure depicted in Fig.~\ref{fig:gate_geometry_basis_potential} (a, b). 
Voltage control of QDs is achieved through three plunger gates   $V_{1-3}$ and two tunnel gates  $W_{1-2}$, while two  screening gates $V_{scr}$ and  four barrier gates $V_b$ are kept grounded at all times.
	We obtain the electrostatic potential $\varphi(x,y,z)$ in this device geometry
	 from a 3D finite-difference Poisson solver \texttt{nextnano++} \cite{Birner_2006} in the empty dot regime, %at $V_1=0.98$ V, $V_2=1.02$ V, $V_3=0.925$ V, and $W_1=W_2=0.75$ V.
	 having chosen a high-bias voltage configuration to demonstrate the applicability of our localization method to highly asymmetric potentials.
	Fig. \ref{fig:gate_geometry_basis_potential}(c) shows the conduction band edge profile $E_c(x,y) = E_g/2 - e\varphi(x,y, z_e)$, where  $E_g = 1.17\ \text{eV}$ is the cryogenic bandgap of Si, at the plane of maximum electron density ($z_e=-1.7$ nm below the
	 Si/SiO$_2$ interface) if electrons were accumulated.
	 Segmentation into three QD regions (white lines and labels) is performed on $E_c(x,y)$ with the random walker algorithm \cite{Grady2006_random_walker} from \texttt{scikit-image} Python package \cite{Walt2014scikitimageimage}. %, with seeds placed at the local minima
	 It is found to yield smoother boundaries, be more robust
	 to voltage-induced variations in the shape of $\varphi(x,y)$, and allow for better-localized orbitals than the compact watershed algorithm \cite{Neubert2014CompactWatershed} from the same library. %despite the latter being able to capture the separatrices of the potential landscape much more closely.
	 $M=33$ lowest-energy eigenstates $\ket{\psi_i(x,y)}$ are found by diagonalizing the Effective mass Hamiltonian $H = -\frac{\hbar^2}{2m^{*} }\Delta_{xy} + E_c(x,y)$ numerically on a rectilinear $x-y$ grid (here, $m^* = 0.19\; m_e$ for electrons in Si). The four lowest-energy eigenstates, shown in Fig. \ref{fig:gate_geometry_basis_potential}(d), exhibit strong delocalization over all three QD regions.
	 $M=33$ is chosen for the $\psi$-space to encompass all $s-$, $p-$, and $d-$shells for the choice of $\lambda_{\mathrm{thr}}=0.97$. The lowest-lying localized $\chi$---orbitals are shown in Fig.~\ref{fig:gate_geometry_basis_potential}(e).

Figure~\ref{fig:tunnel_voltage_sweep} shows the voltage dependences of tunnel couplings between the $s-$like and lowest-lying $p-$like orbitals of all pairs of QDs in the triangular array.
A simultaneous sweep of two tunneling gates $W_1=W_2=W$ up to $W_{\max{}}\approx0.8$ V, when QDs merge, is performed for the cases of no bias between $1^\text{st}$ and $2^\text{nd}$ QDs (markers),  and for the case of $40$ mV bias between them (solid lines). 
The $\psi$-basis size for each data point is chosen to encompass all 
$s-$, $p-$, $d-$ and $f-$shells in each QD, as per Eq. \eqref{eq:minimum_subbasis_main} with $K=4$, with $\lambda_{thr} = 0.9$.
Coupling values involving $p-$orbitals systematically exceed those between $s-$orbitals, which is consistent with the FCI calculations from \cite{Foulk2024TheoryCSDs_many_orbital_CI}.
The exponential voltage dependence of most interorbital tunnel couplings persists over a remarkable range of 5--7 orders of magnitude. This wide dynamic range demonstrates the algorithm's ability to numerically resolve very weak couplings, even in strongly biased systems (in contrast to Ref.~\cite{Foulk2024TheoryCSDs_many_orbital_CI} with reported numerical artifacts in the weak-coupling regime and incompatibility with large interdot bias).
For the mirror-symmetric $1^\text{st}$ and $2^\text{nd}$ gates, the markers and curves in Fig.~\ref{fig:tunnel_voltage_sweep}(a) align almost perfectly in the low-tunneling regime and deviate noticeably only at strong-coupling values of $10s-100s$ of $\mu$eV.
This demonstrates that the weak dependence on symmetric bias, predicted by the WKB framework~\cite{Platt2008WKBAnalysisTunnel},
applies to all inter-orbital tunnel couplings alike.
Remarkably, our method captures finer effects in tunnel
coupling trends for higher orbitals such as the jump in Fig.~\ref{fig:tunnel_voltage_sweep}(b) and the dip in Fig.~\ref{fig:tunnel_voltage_sweep}(c)---especially 
prominent in the biased regime---within the $W\sim 0.7-0.73$ V window.
The voltage increase on the wide tunnel gate $W_2$ causes the third QD to transition from elongation along $x$ at
low $W$ to elongation along $y$ at high $W$. As a result, the lowest-lying $p$-like orbital $\ket{\chi_{1}^{(3)}}$ changes its orientation as illustrated schematically in the inset of Fig.~\ref{fig:tunnel_voltage_sweep}(c).
At $W\sim 0.7-0.73$ V, this $p$-like orbital points approximately
toward the  $2^\text{nd}$ dot but orthogonally to the $1^\text{st}$ dot, which suppresses its tunnel coupling with the $1^\text{st}$ dot and enhances its  coupling with the $2^\text{nd}$ dot.

\begin{figure*} 
	\centering
	\includegraphics[width=0.9\linewidth]{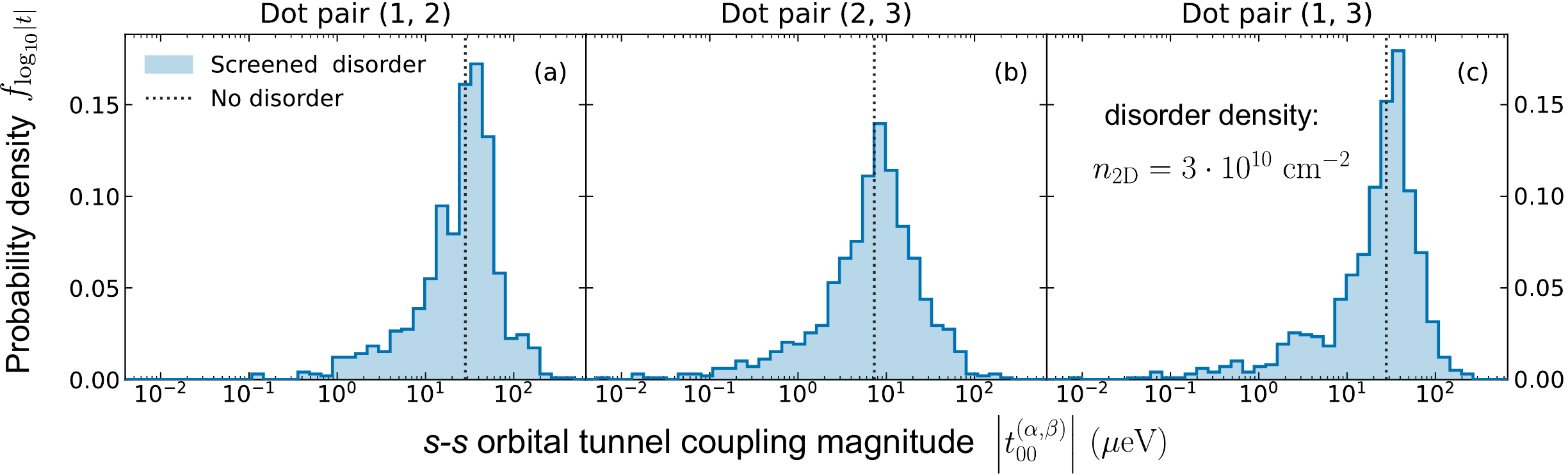}  
	\caption{Estimated probability density functions of pairwise tunnel couplings between $s$-orbitals, $\abs{t^{(\alpha, \beta)}_{00}}$, in the presence of disorder, at the voltage configuration from Fig.~\ref{fig:gate_geometry_basis_potential}. Disorder is simulated as $1000$ random configurations of $n=3$ fixed negative point charges randomly placed in the oxide layers over $100\text{ nm}\times100\text{ nm}$ range in the $x-y$ plane. The spatial distributions of random charges and the incorporation of screening with the method of image charges are detailed in the End Matter.}
	\label{fig:tunnel_disorder_distribution}
\end{figure*}

Realistically, $\psi-$states used as the starting point of our method, are very smooth even in disordered potentials, and the segmentation methods \cite{Grady2006_random_walker, Neubert2014CompactWatershed} efficiently handle weak boundaries. Therefore, the presented localization algorithm is suitable for tunnel coupling probability distribution estimation in the presence of disorder, exemplified in Fig.~\ref{fig:tunnel_disorder_distribution}. Disorder is simulated as fixed point charges randomly placed in the oxide layers, with screening taken into account as discussed in the End Matter.
Additional calculations, not shown, indicate that the distributions of $\log\abs{t^{(\alpha,\beta)}_{ij}}$ are consistently unimodal, with variance and skewness dependent on the sign and number of random charges. An in-depth study of these disorder-induced distributions could be highly beneficial for the experimental identification of the dominant sources of disorder in batches of nominally identical QD devices.

\textit{Implications for many-body calculations}--- 
The presented basis decomposition is very promising for many-electron calculations within Hubbard and configuration-interaction models for several reasons. 
Since each $\chi$---subbasis is maximally localized in its corresponding QD, the products of type $\chi_{i}^{(\alpha)}(\vec{r}) \chi_{j}^{(\beta)}(\vec{r})$ for $\alpha\neq\beta$ can be made negligibly small everywhere in $\mathbb{R}^D$ by choosing values of $\lambda_{thr}$ close to 1. 
This sets a large proportion of the Coulomb matrix elements to zero: 
\begin{equation} 	\bra{\chi_i^{(\alpha)}\chi_j^{(\beta)}}\frac{1}{\abs{\vec{r}_1 - \vec{r}_2} } \ket{\chi_k^{(\gamma)}\chi_l^{(\delta)}} \approx  v^{\alpha\beta}_{ik,jl} \delta_{\alpha\gamma}\delta_{\beta\delta}.
	\label{eq:coulomb_integrals_suppression}	
\end{equation}
The total number of nonzero Coulomb integrals $ v^{\alpha\beta}_{ik,jl}$ is approximately $\frac{1}{16}N^2(M/N)^4 = \frac{1}{16}M^4/N^2$, which is a factor of $\sim 1/2N^2$ smaller than the total number of Coulomb integrals in the full $\psi$-basis of size $M$. Therefore, the localized $\chi$-orbital bases could offer a significant speedup over the Hamiltonian eigenbasis when used for the FCI calculations, especially in arrays with large numbers of QDs.
On the other hand,  the terms of type  $t_{ij}^{(\alpha, \beta)} \left(c_{i}^{(\alpha)} \right)^\dag c_{j}^{(\beta)} $, together with the off-diagonal Coulomb scattering terms, could be added \textit{perturbatively} on top of the Hubbard Hamiltonian.
This could be instrumental in quantifying the tunnel coupling measured at $(1,1)-(0,2)$ or higher-order charge transitions, and building a perturbative framework for exchange calculations in general confinement potentials. 
Conveniently, all the exponential trends of exchange are automatically embedded in the tunnel coupling dependencies, as can be seen even from the result $J\sim t^2/(U\pm \epsilon)$ of the standard ($s-$orbital only) Hubbard model \cite{Wiel2006Semiconductorquantumdots}.

\textit{Conclusions and discussion}---
In conclusion, we have devised a reliable technique of finding localized $s-$, $p-$, $d-,...$ orbital states and their pairwise tunnel couplings in general multi-QD arrays by generalizing the idea of a maximally-localized state, implicit in a two-level-system model for a double-QD, to the full vector space of Hamiltonian eigenstates. 
Thanks to the curated numerical segmentation methods, the procedure is compatible with  geometrically nontrivial and disordered confinement potentials in realistic device geometries. 
By choosing the appropriate Hamiltonian basis size, one either obtains the estimator for the tunnel coupling measured in the single-electron regime, or a promising framework for many-body calculations.
Our method captures the universal nature of the exponential voltage trends in  tunnel coupling between higher-order orbitals and unveils finer orientation-dependent effects.
Although we have focused on a simulated triangular 2D array of Si-MOS QDs throughout this Letter to demonstrate the method,
it is equally applicable to  hole QDs, QD arrays in nanowires 
\cite{Hu2007GeSiheterostructurenanowire, Wang2017CoherentTransportLinear3QD} or 2D materials such as bilayer graphene \cite{Wei2013TuningTunnelingGraphene}, etc., with the appropriate choice of the Hamiltonian operator in Fig.~\ref{fig:flowchart}.

\textit{Acknowledgements}---The authors thank Stephen R. Harrigan for valuable discussions. 
This research was undertaken thanks in part to funding
from the Canada First Research Excellence Fund (Transformative Quantum Technologies) and Canada’s Natural
Sciences and Engineering Research Council (NSERC).
B. K. acknowledges further support from the NSERC
Canada Graduate Scholarship---Doctoral program.

\bibliography{silicon}% Produces the bibliography via BibTeX.

@Article{Volk2019Loadingquantumdot,
  author    = {C. Volk and A. M. J. Zwerver and U. Mukhopadhyay and P. T. Eendebak and C. J. van Diepen and J. P. Dehollain and T. Hensgens and T. Fujita and C. Reichl and W. Wegscheider and L. M. K. Vandersypen},
  journal   = {npj Quantum Information},
  title     = {Loading a quantum-dot based "Qubyte" register},
  year      = {2019},
  issn      = {2056-6387},
  month     = apr,
  number    = {1},
  volume    = {5},
  doi       = {10.1038/s41534-019-0146-y},
  publisher = {Springer Science and Business Media {LLC}},
}

@Article{Burkard_1999,
  author    = {Burkard, Guido and Loss, Daniel and DiVincenzo, David P.},
  journal   = {Physical Review B},
  title     = {Coupled quantum dots as quantum gates},
  year      = {1999},
  issn      = {1095-3795},
  month     = jan,
  number    = {3},
  pages     = {2070--2078},
  volume    = {59},
  doi       = {10.1103/physrevb.59.2070},
  publisher = {American Physical Society (APS)},
}

@Article{Puerto_Gimenez_2007,
  author    = {Puerto Gimenez, Irene and Korkusinski, Marek and Hawrylak, Pawel},
  journal   = {Physical Review B},
  title     = {Linear combination of harmonic orbitals and configuration interaction method for the voltage control of exchange interaction in gated lateral quantum dot networks},
  year      = {2007},
  issn      = {1550-235X},
  month     = aug,
  number    = {7},
  pages     = {075336},
  volume    = {76},
  doi       = {10.1103/physrevb.76.075336},
  publisher = {American Physical Society (APS)},
}

@Article{Nielsen2010Implicationssimultaneousrequirements,
  author    = {Nielsen, Erik and Young, Ralph W. and Muller, Richard P. and Carroll, M. S.},
  journal   = {Physical Review B},
  title     = {Implications of simultaneous requirements for low-noise exchange gates in double quantum dots},
  year      = {2010},
  issn      = {1550-235X},
  month     = aug,
  number    = {7},
  pages     = {075319},
  volume    = {82},
  doi       = {10.1103/physrevb.82.075319},
  publisher = {American Physical Society (APS)},
}

@Article{Pedersen_2007,
  author    = {Pedersen, Jesper and Flindt, Christian and Mortensen, Niels Asger and Jauho, Antti-Pekka},
  journal   = {Physical Review B},
  title     = {Failure of standard approximations of the exchange coupling in nanostructures},
  year      = {2007},
  issn      = {1550-235X},
  month     = sep,
  number    = {12},
  pages     = {125323},
  volume    = {76},
  doi       = {10.1103/physrevb.76.125323},
  publisher = {American Physical Society (APS)},
}

@Article{Birner_2006,
  author    = {Birner, S. and Hackenbuchner, S. and Sabathil, M. and Zandler, G. and Majewski, J.A. and Andlauer, T. and Zibold, T. and Morschl, R. and Trellakis, A. and Vogl, P.},
  journal   = {Acta Physica Polonica A},
  title     = {Modeling of Semiconductor Nanostructures with nextnano3},
  year      = {2006},
  issn      = {1898-794X},
  month     = aug,
  number    = {2},
  pages     = {111--124},
  volume    = {110},
  doi       = {10.12693/aphyspola.110.111},
  publisher = {Institute of Physics, Polish Academy of Sciences},
}

@Article{Philips2022Universalcontrolsix,
  author    = {Stephan G. J. Philips and Mateusz T. M{\k{a}}dzik and Sergey V. Amitonov and Sander L. de Snoo and Maximilian Russ and Nima Kalhor and Christian Volk and William I. L. Lawrie and Delphine Brousse and Larysa Tryputen and Brian Paquelet Wuetz and Amir Sammak and Menno Veldhorst and Giordano Scappucci and Lieven M. K. Vandersypen},
  journal   = {Nature},
  title     = {Universal control of a six-qubit quantum processor in silicon},
  year      = {2022},
  month     = sep,
  number    = {7929},
  pages     = {919--924},
  volume    = {609},
  doi       = {10.1038/s41586-022-05117-x},
  publisher = {Springer Science and Business Media {LLC}},
}

@Article{DiCarlo2004DifferentialChargeSensing,
  author    = {L. DiCarlo and H. J. Lynch and A. C. Johnson and L. I. Childress and K. Crockett and C. M. Marcus and M. P. Hanson and A. C. Gossard},
  journal   = {Physical Review Letters},
  title     = {Differential Charge Sensing and Charge Delocalization in a Tunable Double Quantum Dot},
  year      = {2004},
  month     = jun,
  number    = {22},
  pages     = {226801},
  volume    = {92},
  doi       = {10.1103/physrevlett.92.226801},
  publisher = {American Physical Society ({APS})},
}

@Article{Hsiao2020EfficientOrthogonalControl,
  author    = {T.-K. Hsiao and C.J. van Diepen and U. Mukhopadhyay and C. Reichl and W. Wegscheider and L.M.K. Vandersypen},
  journal   = {Physical Review Applied},
  title     = {Efficient Orthogonal Control of Tunnel Couplings in a Quantum Dot Array},
  year      = {2020},
  month     = may,
  number    = {5},
  pages     = {054018},
  volume    = {13},
  doi       = {10.1103/physrevapplied.13.054018},
  publisher = {American Physical Society ({APS})},
}

@Article{Oosterkamp_1998,
  author    = {Oosterkamp, T. H. and Fujisawa, T. and van der Wiel, W. G. and Ishibashi, K. and Hijman, R. V. and Tarucha, S. and Kouwenhoven, L. P.},
  journal   = {Nature},
  title     = {Microwave spectroscopy of a quantum-dot molecule},
  year      = {1998},
  issn      = {1476-4687},
  month     = oct,
  number    = {6705},
  pages     = {873--876},
  volume    = {395},
  doi       = {10.1038/27617},
  publisher = {Springer Science and Business Media LLC},
}

@Article{Diepen2018Automatedtuninginter,
  author    = {C. J. van Diepen and P. T. Eendebak and B. T. Buijtendorp and U. Mukhopadhyay and T. Fujita and C. Reichl and W. Wegscheider and L. M. K. Vandersypen},
  journal   = {Applied Physics Letters},
  title     = {Automated tuning of inter-dot tunnel coupling in double quantum dots},
  year      = {2018},
  month     = jul,
  number    = {3},
  volume    = {113},
  doi       = {10.1063/1.5031034},
  publisher = {{AIP} Publishing},
}

@Article{DasSarma2011Hubbardmodeldescription,
  author     = {Das Sarma, S. and Wang, Xin and Yang, Shuo},
  journal    = {Physical Review B},
  title      = {Hubbard model description of silicon spin qubits: {Charge} stability diagram and tunnel coupling in {Si} double quantum dots},
  year       = {2011},
  month      = jun,
  number     = {23},
  pages      = {235314},
  volume     = {83},
  doi        = {10.1103/PhysRevB.83.235314},
  publisher  = {American Physical Society},
  shorttitle = {Hubbard model description of silicon spin qubits},
  urldate    = {2023-11-08},
}

@TechReport{Buonacorsi2020Optimizinglateralquantum,
  author      = {Buonacorsi, Brandon and Korkusinski, Marek and Khromets, Bohdan and Baugh, Jonathan},
  institution = {University of Waterloo},
  title       = {Optimizing lateral quantum dot geometries for reduced exchange noise},
  year        = {2020},
  month       = dec,
  note        = {arXiv:2012.10512 [cond-mat, physics:quant-ph] type: article},
  eprint      = {2012.10512},
  eprinttype  = {arxiv},
  school      = {arXiv},
}

@PhdThesis{Platt2008WKBAnalysisTunnel,
  author = {Platt, Edward},
  school = {University of Waterloo},
  title  = {WKB Analysis of Tunnel Coupling in a Simple Model of a Double Quantum Dot},
  year   = {2008},
}

@Article{Anderson2022Highprecisionreal,
  author    = {Anderson, Christopher R. and Gyure, Mark F. and Quinn, Sam and Pan, Andrew and Ross, Richard S. and Kiselev, Andrey A.},
  journal   = {AIP Advances},
  title     = {High-precision real-space simulation of electrostatically confined few-electron states},
  year      = {2022},
  issn      = {2158-3226},
  month     = jun,
  number    = {6},
  volume    = {12},
  date      = {2022-06-28},
  day       = {28},
  doi       = {10.1063/5.0089350},
  publisher = {AIP Publishing},
  source    = {AIP Advances},
}

@Article{Jnane2025Abinitiomodeling,
  author    = {Jnane, Hamza and Benjamin, Simon C.},
  journal   = {Physical Review Applied},
  title     = {Ab initio modeling of quantum dot qubits: Coupling, gate dynamics, and robustness versus charge noise},
  year      = {2025},
  issn      = {2331-7019},
  month     = oct,
  number    = {4},
  volume    = {24},
  doi       = {10.1103/rm43-nh9b},
  publisher = {American Physical Society (APS)},
}

@Article{Huang2018Spindecoherencetwo,
  author    = {Huang, Peihao and Zimmerman, Neil M. and Bryant, Garnett W.},
  journal   = {npj Quantum Information},
  title     = {Spin decoherence in a two-qubit CPHASE gate: the critical role of tunneling noise},
  year      = {2018},
  issn      = {2056-6387},
  month     = nov,
  number    = {1},
  volume    = {4},
  doi       = {10.1038/s41534-018-0112-0},
  publisher = {Springer Science and Business Media LLC},
}

@Article{Bhattacharya1982DoubleMinimumWells,
  author     = {Bhattacharya, S. K. and Rau, A. R. P.},
  journal    = {Physical Review A},
  title      = {Coulomb spectrum in crossed electric and magnetic fields: {Eigenstates} of motion in double-minimum potential wells},
  year       = {1982},
  month      = nov,
  number     = {5},
  pages      = {2315--2321},
  volume     = {26},
  doi        = {10.1103/PhysRevA.26.2315},
  publisher  = {American Physical Society},
  shorttitle = {Coulomb spectrum in crossed electric and magnetic fields},
  url        = {https://link.aps.org/doi/10.1103/PhysRevA.26.2315},
  urldate    = {2026-05-17},
}

@Article{Klos2018Calculationtunnelcouplings,
  author    = {Klos, Jan and Hassler, Fabian and Cerfontaine, Pascal and Bluhm, Hendrik and Schreiber, Lars R.},
  journal   = {Physical Review B},
  title     = {Calculation of tunnel couplings in open gate-defined disordered quantum dot systems},
  year      = {2018},
  issn      = {2469-9969},
  month     = Oct,
  number    = {15},
  pages     = {155320},
  volume    = {98},
  doi       = {10.1103/physrevb.98.155320},
  publisher = {American Physical Society (APS)},
}

@Article{Foulk2024TheoryCSDs_many_orbital_CI,
  author    = {Foulk, Nathan L. and Das Sarma, Sankar},
  journal   = {Physical Review B},
  title     = {Theory of charge stability diagrams in coupled quantum dot qubits},
  year      = {2024},
  issn      = {2469-9969},
  month     = Nov,
  number    = {20},
  pages     = {205428},
  volume    = {110},
  doi       = {10.1103/physrevb.110.205428},
  publisher = {American Physical Society (APS)},
}

@Article{Yang2011GenericHubbardmodel,
  author    = {Yang, Shuo and Wang, Xin and Das Sarma, S.},
  journal   = {Physical Review B},
  title     = {Generic Hubbard model description of semiconductor quantum-dot spin qubits},
  year      = {2011},
  issn      = {1550-235X},
  month     = Apr,
  number    = {16},
  pages     = {161301},
  volume    = {83},
  doi       = {10.1103/physrevb.83.161301},
  publisher = {American Physical Society (APS)},
}

@Article{Hatano2005SingleElectronDelocalization,
  author    = {Hatano, T. and Stopa, M. and Tarucha, S.},
  journal   = {Science},
  title     = {Single-Electron Delocalization in Hybrid Vertical-Lateral Double Quantum Dots},
  year      = {2005},
  issn      = {1095-9203},
  month     = jul,
  number    = {5732},
  pages     = {268--271},
  volume    = {309},
  doi       = {10.1126/science.1111205},
  publisher = {American Association for the Advancement of Science (AAAS)},
}

@Article{Huettel2005Directcontroltunnel,
  author    = {Hüttel, A. K. and Ludwig, S. and Lorenz, H. and Eberl, K. and Kotthaus, J. P.},
  journal   = {Physical Review B},
  title     = {Direct control of the tunnel splitting in a one-electron double quantum dot},
  year      = {2005},
  issn      = {1550-235X},
  month     = Aug,
  number    = {8},
  pages     = {081310},
  volume    = {72},
  doi       = {10.1103/physrevb.72.081310},
  publisher = {American Physical Society (APS)},
}

@Article{PioroLadriere2005ChargeSensingArtificialH2+molecule,
  author    = {Pioro-Ladrière, M. and Abolfath, M. R. and Zawadzki, P. and Lapointe, J. and Studenikin, S. A. and Sachrajda, A. S. and Hawrylak, P.},
  journal   = {Physical Review B},
  title     = {Charge sensing of an artificialH2+molecule in lateral quantum dots},
  year      = {2005},
  issn      = {1550-235X},
  month     = sep,
  number    = {12},
  pages     = {125307},
  volume    = {72},
  doi       = {10.1103/physrevb.72.125307},
  publisher = {American Physical Society (APS)},
}

@Article{Simon1985FleaElephant,
  author    = {Simon, Barry},
  journal   = {Journal of Functional Analysis},
  title     = {Semiclassical analysis of low lying eigenvalues. IV. The flea on the elephant},
  year      = {1985},
  issn      = {0022-1236},
  month     = Aug,
  number    = {1},
  pages     = {123--136},
  volume    = {63},
  doi       = {10.1016/0022-1236(85)90101-6},
  publisher = {Elsevier BV},
}

@Article{Recher2009Boundstatesmagnetic,
  author    = {Recher, Patrik and Nilsson, Johan and Burkard, Guido and Trauzettel, Björn},
  journal   = {Physical Review B},
  title     = {Bound states and magnetic field induced valley splitting in gate-tunable graphene quantum dots},
  year      = {2009},
  month     = feb,
  number    = {8},
  pages     = {085407},
  volume    = {79},
  doi       = {10.1103/PhysRevB.79.085407},
  publisher = {American Physical Society},
  url       = {https://link.aps.org/doi/10.1103/PhysRevB.79.085407},
  urldate   = {2026-07-01},
}

@Article{Wei2013TuningTunnelingGraphene,
  author    = {Wei, Da and Li, Hai-Ou and Cao, Gang and Luo, Gang and Zheng, Zhi-Xiong and Tu, Tao and Xiao, Ming and Guo, Guang-Can and Jiang, Hong-Wen and Guo, Guo-Ping},
  journal   = {Scientific Reports},
  title     = {Tuning inter-dot tunnel coupling of an etched graphene double quantum dot by adjacent metal gates},
  year      = {2013},
  issn      = {2045-2322},
  month     = nov,
  number    = {1},
  pages     = {3175},
  volume    = {3},
  copyright = {2013 The Author(s)},
  doi       = {10.1038/srep03175},
  publisher = {Nature Publishing Group},
  url       = {https://www.nature.com/articles/srep03175},
  urldate   = {2026-07-01},
}

@Article{Beenakker2008AndreevreflectionGraphene,
  author     = {Beenakker, C. W. J.},
  journal    = {Reviews of Modern Physics},
  title      = {Colloquium: {Andreev} reflection and {Klein} tunneling in graphene},
  year       = {2008},
  month      = oct,
  number     = {4},
  pages      = {1337--1354},
  volume     = {80},
  doi        = {10.1103/RevModPhys.80.1337},
  publisher  = {American Physical Society},
  shorttitle = {Colloquium},
  url        = {https://link.aps.org/doi/10.1103/RevModPhys.80.1337},
  urldate    = {2026-07-01},
}

@Article{Hu2007GeSiheterostructurenanowire,
  author    = {Hu, Yongjie and Churchill, Hugh O. H. and Reilly, David J. and Xiang, Jie and Lieber, Charles M. and Marcus, Charles M.},
  journal   = {Nature Nanotechnology},
  title     = {A {Ge}/{Si} heterostructure nanowire-based double quantum dot with integrated charge sensor},
  year      = {2007},
  issn      = {1748-3395},
  month     = oct,
  number    = {10},
  pages     = {622--625},
  volume    = {2},
  copyright = {2007 Springer Nature Limited},
  doi       = {10.1038/nnano.2007.302},
  publisher = {Nature Publishing Group},
  url       = {https://www.nature.com/articles/nnano.2007.302},
  urldate   = {2026-07-01},
}

@Article{Braakman2013LongdistanceCoherentCoupling,
  author    = {Braakman, F. R. and Barthelemy, P. and Reichl, C. and Wegscheider, W. and Vandersypen, L. M. K.},
  journal   = {Nature Nanotechnology},
  title     = {Long-distance coherent coupling in a quantum dot array},
  year      = {2013},
  issn      = {1748-3395},
  month     = jun,
  number    = {6},
  pages     = {432--437},
  volume    = {8},
  copyright = {2013 Springer Nature Limited},
  doi       = {10.1038/nnano.2013.67},
  publisher = {Nature Publishing Group},
  url       = {https://www.nature.com/articles/nnano.2013.67},
  urldate   = {2026-08-09},
}

@Article{Noiri2017triangular3QD,
  author    = {Noiri, A and Kawasaki, K and Otsuka, T and Nakajima, T and Yoneda, J and Amaha, S and Delbecq, M R and Takeda, K and Allison, G and Ludwig, A and Wieck, A D and Tarucha, S},
  journal   = {Semiconductor Science and Technology},
  title     = {A triangular triple quantum dot with tunable tunnel couplings},
  year      = {2017},
  issn      = {0268-1242},
  month     = jul,
  number    = {8},
  pages     = {084004},
  volume    = {32},
  doi       = {10.1088/1361-6641/aa7596},
  language  = {en},
  publisher = {IOP Publishing},
  url       = {https://doi.org/10.1088/1361-6641/aa7596},
  urldate   = {2026-08-11},
}

@Article{Mukhopadhyay2018_2x2,
  author   = {Mukhopadhyay, Uditendu and Dehollain, Juan Pablo and Reichl, Christian and Wegscheider, Werner and Vandersypen, Lieven M. K.},
  journal  = {Applied Physics Letters},
  title    = {A 2 × 2 quantum dot array with controllable inter-dot tunnel couplings},
  year     = {2018},
  issn     = {0003-6951},
  month    = apr,
  number   = {18},
  pages    = {183505},
  volume   = {112},
  doi      = {10.1063/1.5025928},
  url      = {https://doi.org/10.1063/1.5025928},
  urldate  = {2026-08-11},
}

@Article{Wang2024_Si_2x2,
  author   = {Wang, Ning and Kang, Jia-Min and Lu, Wen-Long and Wang, Shao-Min and Wang, You-Jia and Li, Hai-Ou and Cao, Gang and Wang, Bao-Chuan and Guo, Guo-Ping},
  journal  = {Nano Letters},
  title    = {Highly {Tunable} {2D} {Silicon} {Quantum} {Dot} {Array} with {Coupling} beyond {Nearest} {Neighbors}},
  year     = {2024},
  issn     = {1530-6984},
  month    = oct,
  number   = {42},
  pages    = {13126--13133},
  volume   = {24},
  doi      = {10.1021/acs.nanolett.4c02345},
  url      = {https://doi.org/10.1021/acs.nanolett.4c02345},
  urldate  = {2026-08-11},
}

@Article{Zhao2022MeasurementTunnelCoupling,
  author    = {Zhao, Xinyu and Hu, Xuedong},
  journal   = {Physical Review Applied},
  title     = {Measurement of Tunnel Coupling in a {S}i Double Quantum dot Based on Charge Sensing},
  year      = {2022},
  issn      = {2331-7019},
  month     = jun,
  number    = {6},
  pages     = {064043},
  volume    = {17},
  doi       = {10.1103/physrevapplied.17.064043},
  publisher = {American Physical Society (APS)},
  url       = {https://link.aps.org/doi/10.1103/PhysRevApplied.17.064043},
  urldate   = {2025-05-10},
}

@Article{Zwolak2018QFlowlitedataset,
  author     = {Zwolak, Justyna P. and Kalantre, Sandesh S. and Wu, Xingyao and Ragole, Stephen and Taylor, Jacob M.},
  journal    = {PLOS ONE},
  title      = {{QFlow} lite dataset: {A} machine-learning approach to the charge states in quantum dot experiments},
  year       = {2018},
  issn       = {1932-6203},
  month      = oct,
  number     = {10},
  pages      = {e0205844},
  volume     = {13},
  doi        = {10.1371/journal.pone.0205844},
  language   = {en},
  publisher  = {Public Library of Science},
  shorttitle = {{QFlow} lite dataset},
  url        = {https://journals.plos.org/plosone/article?id=10.1371/journal.pone.0205844},
  urldate    = {2026-08-11},
}

@Article{Wang2017CoherentTransportLinear3QD,
  author   = {Wang, Ji-Yin and Huang, Shaoyun and Huang, Guang-Yao and Pan, Dong and Zhao, Jianhua and Xu, H. Q.},
  journal  = {Nano Letters},
  title    = {Coherent {Transport} in a {Linear} {Triple} {Quantum} {Dot} {Made} from a {Pure}-{Phase} {InAs} {Nanowire}},
  year     = {2017},
  issn     = {1530-6984},
  month    = jun,
  number   = {7},
  pages    = {4158--4164},
  volume   = {17},
  doi      = {10.1021/acs.nanolett.7b00927},
  url      = {https://doi.org/10.1021/acs.nanolett.7b00927},
  urldate  = {2026-08-11},
}

@Article{Brandes2002Adiabatictransferelectrons,
  author    = {Brandes, T. and Vorrath, T.},
  journal   = {Physical Review B},
  title     = {Adiabatic transfer of electrons in coupled quantum dots},
  year      = {2002},
  month     = aug,
  number    = {7},
  pages     = {075341},
  volume    = {66},
  doi       = {10.1103/PhysRevB.66.075341},
  publisher = {American Physical Society},
  url       = {https://link.aps.org/doi/10.1103/PhysRevB.66.075341},
  urldate   = {2026-08-11},
}

@Article{DeSmet2025Highfidelitysingle,
  author    = {De Smet, Maxim and Matsumoto, Yuta and Zwerver, Anne-Marije J. and Tryputen, Larysa and de Snoo, Sander L. and Amitonov, Sergey V. and Katiraee-Far, Sam R. and Sammak, Amir and Samkharadze, Nodar and G{\"u}l, {\"O}nder and Wasserman, Rick N. M. and Greplov{\'a}, Eli{\v{s}}ka and Rimbach-Russ, Maximilian and Scappucci, Giordano and Vandersypen, Lieven M. K.},
  journal   = {Nature Nanotechnology},
  title     = {High-fidelity single-spin shuttling in silicon},
  year      = {2025},
  issn      = {1748-3395},
  month     = jun,
  number    = {7},
  pages     = {866--872},
  volume    = {20},
  doi       = {10.1038/s41565-025-01920-5},
  publisher = {Springer Science and Business Media LLC},
}

@Article{Madzik2025Operatingtwoexchange,
  author    = {Madzik, Mateusz T. and Luthi, Florian and Guerreschi, Gian Giacomo and Mohiyaddin, Fahd A. and Borjans, Felix and Chadwick, Jason D. and Curry, Matthew J. and Ziegler, Joshua and Atanasov, Sarah and Bavdaz, Peter L. and Connors, Elliot J. and Corrigan, J. and Ercan, H. Ekmel and Flory, Robert and George, Hubert C. and Harpt, Benjamin and Henry, Eric and Islam, Mohammad M. and Khammassi, Nader and Keith, Daniel and Lampert, Lester F. and Mladenov, Todor M. and Morris, Randy W. and Nethwewala, Aditi and Neyens, Samuel and Otten, René and Osuna Ibarra, Linda P. and Patra, Bishnu and Pillarisetty, Ravi and Premaratne, Shavindra and Ramsey, Mick and Risinger, Andrew and Rooney, John D. and Savytskyy, Rostyslav and Watson, Thomas F. and Zietz, Otto K. and Matsuura, Anne Y. and Pellerano, Stefano and Bishop, Nathaniel C. and Roberts, Jeanette and Clarke, James S.},
  journal   = {Nature},
  title     = {Operating two exchange-only qubits in parallel},
  year      = {2025},
  issn      = {1476-4687},
  month     = nov,
  number    = {8091},
  pages     = {870--875},
  volume    = {647},
  copyright = {2025 The Author(s)},
  doi       = {10.1038/s41586-025-09767-5},
  language  = {en},
  publisher = {Nature Publishing Group},
  url       = {https://www.nature.com/articles/s41586-025-09767-5},
  urldate   = {2026-08-11},
}

@Article{Kandel2021Adiabaticquantumstate,
  author    = {Kandel, Yadav P. and Qiao, Haifeng and Fallahi, Saeed and Gardner, Geoffrey C. and Manfra, Michael J. and Nichol, John M.},
  journal   = {Nature Communications},
  title     = {Adiabatic quantum state transfer in a semiconductor quantum-dot spin chain},
  year      = {2021},
  issn      = {2041-1723},
  month     = apr,
  number    = {1},
  pages     = {2156},
  volume    = {12},
  copyright = {2021 The Author(s)},
  doi       = {10.1038/s41467-021-22416-5},
  language  = {en},
  publisher = {Nature Publishing Group},
  url       = {https://www.nature.com/articles/s41467-021-22416-5},
  urldate   = {2026-08-12},
}

@Article{Borjans2021Intervalley_Tunnel_Coupling,
  author    = {Borjans, F. and Zhang, X. and Mi, X. and Cheng, G. and Yao, N. and Jackson, C.A.C. and Edge, L.F. and Petta, J.R.},
  journal   = {PRX Quantum},
  title     = {Probing the {Variation} of the {Intervalley} {Tunnel} {Coupling} in a {Silicon} {Triple} {Quantum} {Dot}},
  year      = {2021},
  month     = apr,
  number    = {2},
  pages     = {020309},
  volume    = {2},
  doi       = {10.1103/PRXQuantum.2.020309},
  publisher = {American Physical Society},
  url       = {https://link.aps.org/doi/10.1103/PRXQuantum.2.020309},
  urldate   = {2026-08-12},
}

@InProceedings{Neubert2014CompactWatershed,
  author     = {Neubert, Peer and Protzel, Peter},
  title      = {Compact {Watershed} and {Preemptive} {SLIC}: {On} {Improving} {Trade}-offs of {Superpixel} {Segmentation} {Algorithms}},
  year       = {2014},
  month      = aug,
  note       = {ISSN: 1051-4651},
  pages      = {996--1001},
  doi        = {10.1109/ICPR.2014.181},
  issn       = {1051-4651},
  shorttitle = {Compact {Watershed} and {Preemptive} {SLIC}},
  url        = {https://ieeexplore.ieee.org/abstract/document/6976891},
  urldate    = {2026-08-17},
}

@Article{Walt2014scikitimageimage,
  author     = {Walt, Stéfan van der and Schönberger, Johannes L. and Nunez-Iglesias, Juan and Boulogne, François and Warner, Joshua D. and Yager, Neil and Gouillart, Emmanuelle and Yu, Tony},
  journal    = {PeerJ},
  title      = {scikit-image: image processing in {Python}},
  year       = {2014},
  issn       = {2167-8359},
  month      = jun,
  pages      = {e453},
  volume     = {2},
  doi        = {10.7717/peerj.453},
  language   = {en},
  publisher  = {PeerJ Inc.},
  shorttitle = {scikit-image},
  url        = {https://peerj.com/articles/453},
  urldate    = {2026-08-17},
}

@Article{Grady2006_random_walker,
  author   = {Grady, L.},
  journal  = {IEEE Transactions on Pattern Analysis and Machine Intelligence},
  title    = {Random Walks for Image Segmentation},
  year     = {2006},
  number   = {11},
  pages    = {1768-1783},
  volume   = {28},
  doi      = {10.1109/TPAMI.2006.233},
}

@Article{Leon2020spdf_electrons,
  author    = {Leon, R. C. C. and Yang, C. H. and Hwang, J. C. C. and Lemyre, J. Camirand and Tanttu, T. and Huang, W. and Chan, K. W. and Tan, K. Y. and Hudson, F. E. and Itoh, K. M. and Morello, A. and Laucht, A. and Pioro-Ladrière, M. and Saraiva, A. and Dzurak, A. S.},
  journal   = {Nature Communications},
  title     = {Coherent spin control of s-, p-, d- and f-electrons in a silicon quantum dot},
  year      = {2020},
  issn      = {2041-1723},
  month     = feb,
  number    = {1},
  pages     = {797},
  volume    = {11},
  copyright = {2020 The Author(s)},
  doi       = {10.1038/s41467-019-14053-w},
  language  = {en},
  publisher = {Nature Publishing Group},
  url       = {https://www.nature.com/articles/s41467-019-14053-w},
  urldate   = {2026-08-17},
}

@Article{Jaworowski2017MacroscopicSingletTriplet,
  author    = {Jaworowski, Blazej and Rogers, Nick and Grabowski, Marek and Hawrylak, Pawel},
  journal   = {Scientific Reports},
  title     = {Macroscopic {Singlet}-{Triplet} {Qubit} in {Synthetic} {Spin}-{One} {Chain} in {Semiconductor} {Nanowires}},
  year      = {2017},
  issn      = {2045-2322},
  month     = jul,
  number    = {1},
  pages     = {5529},
  volume    = {7},
  copyright = {2017 The Author(s)},
  doi       = {10.1038/s41598-017-05655-9},
  language  = {en},
  publisher = {Nature Publishing Group},
  url       = {https://www.nature.com/articles/s41598-017-05655-9},
  urldate   = {2026-08-17},
}

@Article{Cockins2010EnergylevelsQDs,
  author    = {Cockins, Lynda and Miyahara, Yoichi and Bennett, Steven D. and Clerk, Aashish A. and Studenikin, Sergei and Poole, Philip and Sachrajda, Andrew and Grutter, Peter},
  journal   = {Proceedings of the National Academy of Sciences},
  title     = {Energy levels of few-electron quantum dots imaged and characterized by atomic force microscopy},
  year      = {2010},
  issn      = {1091-6490},
  month     = May,
  number    = {21},
  pages     = {9496--9501},
  volume    = {107},
  doi       = {10.1073/pnas.0912716107},
  publisher = {National Academy of Sciences},
}

@Article{Vachon_2009_QD_spectrum,
  author    = {Vachon, M. and Raymond, S. and Babinski, A. and Lapointe, J. and Wasilewski, Z. and Potemski, M.},
  journal   = {Physical Review B},
  title     = {Energy shell structure of a single InAs/GaAs quantum dot with a spin-orbit interaction},
  year      = {2009},
  issn      = {1550-235X},
  month     = Apr,
  number    = {16},
  pages     = {165427},
  volume    = {79},
  doi       = {10.1103/physrevb.79.165427},
  publisher = {American Physical Society (APS)},
}

@Article{Chan2018validitymicroscopiccalculations,
  author   = {Chan, GuoXuan and Wang, Xin},
  journal  = {Science China Physics, Mechanics \& Astronomy},
  title    = {On the validity of microscopic calculations of double-quantum-dot spin qubits based on {Fock}-{Darwin} states},
  year     = {2018},
  issn     = {1869-1927},
  month    = jan,
  number   = {4},
  pages    = {040313},
  volume   = {61},
  doi      = {10.1007/s11433-017-9145-6},
  language = {en},
  url      = {https://doi.org/10.1007/s11433-017-9145-6},
  urldate  = {2026-08-25},
}

@Article{Wilk2001Highkgate,
  author     = {Wilk, G. D. and Wallace, R. M. and Anthony, J. M.},
  journal    = {Journal of Applied Physics},
  title      = {High-{$\kappa$} gate dielectrics: {Current} status and materials properties considerations},
  year       = {2001},
  issn       = {0021-8979},
  month      = may,
  number     = {10},
  pages      = {5243--5275},
  volume     = {89},
  doi        = {10.1063/1.1361065},
  shorttitle = {High-{$\kappa$} gate dielectrics},
  url        = {https://doi.org/10.1063/1.1361065},
  urldate    = {2026-08-25},
}

@Article{Wiel2006Semiconductorquantumdots,
  author  = {Wiel, W G van der and Stopa, M and Kodera, T and Hatano, T and Tarucha, S},
  journal = {New Journal of Physics},
  title   = {Semiconductor quantum dots for electron spin qubits},
  year    = {2006},
  issn    = {1367-2630},
  month   = feb,
  number  = {2},
  pages   = {28--28},
  volume  = {8},
  doi     = {10.1088/1367-2630/8/2/028},
  url     = {https://iopscience.iop.org/article/10.1088/1367-2630/8/2/028},
  urldate = {2026-08-27},
}
%\bibliography{../../../Insync/hromecb@gmail.com/GOOGLE\string~1/Articles/silicon}

\onecolumngrid
\section*{End Matter}
\twocolumngrid

\textit{Derivation of the localization procedure}---
We consider a system of $N$ quantum dots (QDs) in an arbitrary external confinement potential landscape in $D$ spatial dimensions.	
Let each $\alpha^\text{th}$ QD be associated with a unique spatial domain $Q^{(\alpha)}\in{\mathrm{R}}^{D}$ such that 
%$\bigcup_{\alpha=1}^{N} Q^{(\alpha)} = \mathbb{R}^n$, and 
$Q^{(\alpha)} \cap{Q^{(\beta)} }=\emptyset$ for $\alpha\neq \beta$. 
It is natural to identify a single-electron state $\ket{\xi^{(\alpha)}(\vec{r})}$ as \textit{localized within $Q^{(\alpha)}$} if its total probability on this domain is maximized, or, likewise, as one  \textit{localized outside $Q^{(\alpha)}$} if this total probability is minimized.  Such  states, therefore, extremize the following functional with one Lagrange multiplier due to  normalization $\braket{\xi^{(\alpha)}}{\xi^{(\alpha)}} =1 $:
\begin{equation}
	\mathcal{L}[\xi^{(\alpha)}] = \bra{\xi^{(\alpha)}} \mathbbm{1}_{{Q}^{(\alpha)}}  \ket{\xi^{(\alpha)}} - \lambda^{(\alpha)} \braket{\xi^{(\alpha)}}{\xi^{(\alpha)}},  % \rightarrow \mathrm{extr}
	\label{eq:lagrangian}
\end{equation}
 where  $\mathbbm{1}_{{Q}^{(\alpha)}}(\vec{r})$ is the indicator function on $Q^{(\alpha)}$. 
We expand $\ket{\xi^{(\alpha)}}$ in terms of the $M$ lowest-energy Hamiltonian eigenstates with the vector of coefficients $\vec{C}^{(\alpha)}$:
\begin{equation}
	\ket{\xi^{(\alpha)}} = \sum_{i=1}^{M}\left(\vec{C}^{(\alpha)}\right)_i \ket{\psi_i}, 
	\label{eq:bound_states}
\end{equation}
and substitute 
into \eqref{eq:lagrangian}. The extremum condition $\partial \mathcal{L} /  \partial \vec{C}^{(\alpha)} =\vec{0}$ leads to the eigenvalue problem: 
\begin{equation}
	\mathcal{O}^{(\alpha)} \vec{C}^{(\alpha)} = \lambda^{(\alpha)} \vec{C}^{(\alpha)},  \quad 
	\mathcal{O}^{(\alpha)}_{ij} = \bra{\psi_i} \mathbbm{1}_{{Q}^{(\alpha)}} \ket{\psi_j}. 
	\label{eq:eigenvalue_problem}
\end{equation}
For the orbital state $\ket{\xi_k^{(\alpha)}}$ represented by the $k^\mathrm{th}$ eigenvector $\vec{C}^{(\alpha)}_k$, the eigenvalue $\lambda_k^{(\alpha)}$ is the measure of its total probability within the $\alpha^{th}$ QD: $\lambda_k^{(\alpha)} = \bra{\xi_k^{(\alpha)}} \mathbbm{1}_{{Q}^{(\alpha)}} \ket{\xi_k^{(\alpha)}} \equiv \int_{Q^{\alpha}}\dd{\vec{r}} \abs{\xi_k^{(\alpha)}(\vec{r})}^2 $, which, by construction, is a real quantity within the range of $[0,1]$. 
Therefore, one can choose a threshold value of localization $\lambda_{thr}$  close to 1 to obtain the orthogonal subbasis $\Xi^{(\alpha)}\equiv \left\{\ket{\xi_k^{(\alpha)}} : \lambda_k^{(\alpha)} \geq \lambda_{thr}\right\}$  localized within the $\alpha^{th}$ QD. 
The orthogonal complement of this subbasis, $ \widetilde{\Xi}^{(\alpha)}\equiv\left\{\ket{\xi_k^{(\alpha)}} : \lambda_k^{(\alpha)} < \lambda_{thr}\right\}$, is spread over other QD regions and can be used to perform localization there. 
Fig.~\ref{fig:flowchart} encapsulates this idea into a sequential procedure. At first, one uses the full $\psi$-space to find the bases localized within and outside the first QD region  $Q^{(1)}$. 
For each next $\alpha^\text{th}$ QD, one solves the eigenvalue problem analogous to Eq.~\eqref{eq:eigenvalue_problem} on the subset  of $\ket{\xi^{(\alpha-1)}_i} \in \widetilde{\Xi}^{(\alpha-1)}$:
\begin{equation}
	\begin{split}
	\mathcal{O}^{(\alpha)} \vec{C}^{(\alpha)} = \lambda^{(\alpha)} \vec{C}^{(\alpha)},  \quad
	\mathcal{O}^{(\alpha)}_{ij} = \bra{\xi_i^{(\alpha-1)}} \mathbbm{1}_{{Q}^{(\alpha)}} \ket{\xi_j^{(\alpha-1)}}, 
	\label{eq:eigenvalue_problem_general}
	\end{split}
\end{equation}
 Each step produces two orthogonal subspaces: $\mathrm{Span}\left(\Xi^{(\alpha)}\right)$ and $\mathrm{Span}\left(\widetilde{\Xi}^{(\alpha)}\right)$, and the latter is split into two subspaces at each next step, and so on. Typically, there is a non-empty set $\widetilde{\Xi}^{(N)}$ remaining after the final step of the procedure. It corresponds to the quasi-degenerate high-energy shells whose vector space is only partially spanned by $M$ lowest-lying $\psi$---states.

 We now independently project the Hamiltonian onto each of the mutually orthogonal subspaces $\Xi^{(\alpha)}$.The diagonalization yields the stationary localized orbitals $\ket{\chi_k^{(\alpha)}}$ with level energies $\epsilon_k^{(\alpha)}$ for each $\alpha^\text{th}$ QD:
\begin{equation}
	\mathcal{H}^{(\alpha)}  \ket{\chi_k^{(\alpha)}} = \epsilon_k^{(\alpha)}  \ket{\chi_k^{(\alpha)}}, \quad  \mathcal{H}_{ij}^{(\alpha)}  = \bra{\xi_i^{(\alpha)}} H \ket{\xi_j^{(\alpha)}},
	\label{eq:localized_orbitals}
\end{equation}
Since each $ \ket{\chi_k^{(\alpha)}}\in \mathrm{Span}\left({\Xi} \right)$, the variational theorem guarantees that these states are localized within $\alpha^\text{th}$ QD:
\begin{equation}
\begin{split}
    &\forall i, k:  \ket{\chi_k^{(\alpha)}},  \ket{\xi_i^{(\alpha)}}\in \mathrm{Span}\left({\Xi}^{(\alpha)} \right): \\
	 &\bra{\chi_k^{(\alpha)}} \mathbbm{1}_{{Q}^{(\alpha)}} \ket{\chi_k^{(\alpha)}} \geq \bra{\xi_i^{(\alpha)}} \mathbbm{1}_{{Q}^{(\alpha)}} \ket{\xi_i^{(\alpha)}}\geq \lambda_{thr}.
\end{split}
\label{eq:variational_theorem}
\end{equation}
Projected Hamiltonians from Eq.~\eqref{eq:localized_orbitals} are most convenient to diagonalize in the $\{\ket{\psi_i}\}$-basis representation, where the full Hamiltonian is diagonal.
Using Eq.\eqref{eq:bound_states}, we obtain the matrix form of $\mathcal{H}^{(\alpha)}$:
\begin{equation}
	\mathcal{H}^{(\alpha)} %_{ij} = \sum_{l=1}^{N} E_l \left(\vec{C}_i^{(\alpha)}\right)_{l} \left(\vec{C}_j^{(\alpha)}\right)_{l}  
	= 
	\left[
		\vec{C}_1^{(\alpha)} \cdots  \vec{C}_{M_{\alpha}}^{(\alpha)}
	\right]^\dag \hat{E}
	\left[	\vec{C}_1^{(\alpha)}  \cdots  \vec{C}_{M_{\alpha}}^{(\alpha)}
	\right],
\end{equation}
where $\hat{E} = 
	\mathrm{diag}\left[E_1 \cdots E_M\right] 	$, and $M_{\alpha}$ is the cardinality of $\Xi^{(\alpha)}$.
The tunnel coupling parameters are then given as the matrix elements of the full system Hamiltonian between the corresponding localized orbitals:  
\begin{equation}
	t_{ij}^{\alpha\beta} = \bra{\chi_i^{(\alpha)}} H \ket{\chi_j^{(\beta)}}. 
\end{equation}
As each $\chi-$state is defined up to a global phase, $t_{ij}^{\alpha\beta}$ is itself defined up to the global phases of $\ket{\chi_i^{(\alpha)}}$ and $\ket{\chi_j^{(\beta)}}$ states. Hence, absolute values $\abs{t_{ij}^{\alpha\beta}}$ are plotted in the main text. 
In the $\{\ket{\psi_i}\}$-basis representation, the expression for each tunnel coupling matrix $t^{\alpha\beta}=\left(t^{\beta\alpha}\right)^\dag$ of size $[M_\alpha \times M_\beta]$, which contains all pairwise couplings between the orbitals of the $\alpha^\text{th}$ and the  $\beta^\text{th}$ QDs, simplifies to the following: 
\begin{equation}
	t^{\alpha\beta}
	= \left[\ket{\chi_1^{(\alpha)}}  \cdots \ket{\chi_{M_{\alpha}}^{(\alpha)}}\right]^\dag
		\hat{E}
		\left[
		\ket{\chi_1^{(\beta)}}  \cdots   \ket{\chi_{M_{\beta}}^{(\beta)}}
	\right]
	\label{eq:tunnel_coupling_matrix}
\end{equation}

An important property of the localized $\chi$-orbitals is that the tunnel couplings are nonzero only  between the states in distinct dots $\alpha\neq \beta$. To prove this, we associate the mutually orthogonal subbases $\left\{ \mathrm\Xi^{(1)}, \ldots, \Xi^{(N)}, \widetilde{\Xi}^{(N)}\right\}$ with a complete set of orthogonal projectors $\left\{P_{a}\right\}$, where $a = 1, \ldots, N+1$.
For different orbitals $i\neq j$ in the same $\alpha^\text{th}$ QD,
inserting the resolution of identity into Eq. (8) twice yields zero off-diagonal tunnel coupling terms: 
\begin{equation}
	t^{(\alpha\alpha)}_{ij} = \sum_{a,b=1}^{N+1} \bra{\chi_i^{(\alpha)}} P_{a} H P_{b} \ket{\chi_j^{(\alpha)}}  = \epsilon_i \delta_{ij}.
\end{equation}  
The last equality follows from the fact that $P_b\ket{\chi_j^{(\alpha)}} \equiv 0$ for $b\neq \alpha$, and that $P_\alpha H P_\alpha\equiv \mathcal{H}^{(\alpha)}$, for which $\ket{\chi_{i}^{(\alpha)}}$ are eigenvectors, according to  Eq.~\eqref{eq:localized_orbitals}. 

The obtained localized orbitals and tunnel couplings between them ultimately depend on $\lambda_{thr}$, the number of $\psi$-states $M$, and  the choice of the segmentation procedure. It is recommended to take $\lambda_{thr}\approx 0.9$ (not immediately close to 1) to account for the imperfections of the segmentation routine, 
especially for potential landscapes with shallow tunnel barriers (and thus strong $\psi$---basis delocalization).
The choice of $M$ is discussed below.  

%
%IT IS NOT EVIDENT!
% It is evident from Eq.~\eqref{eq:tunnel_coupling_matrix} that $t^{\alpha\alpha} \equiv \hat{E}$ is a diagonal matrix, since the orbitals localized within the same $\alpha^\text{th}$ $QD$ are mutually orthogonal. This indicates that with this basis choice, the only non-zero tunnel couplings are those between orbitals localized in different QDs.

% \begin{widetext}
% 	\begin{equation}
% 		t^{\alpha\beta}=\left(t^{\beta\alpha}\right)^\dag
% 		= 
% 		\begin{bmatrix}
% 			\ket{\chi_1^{(\alpha)}} & \cdots & \ket{\chi_{M_{\alpha}}^{(\alpha)}}
% 		\end{bmatrix}^\dag
% 		\mathrm{diag}\left[E_1, \ldots, E_N\right] 	\begin{bmatrix}
% 			\ket{\chi_1^{(\beta)}} & \cdots & \ket{\chi_{M_{\beta}}^{(\beta)}}
% 		\end{bmatrix}
% 		\label{eq:tunnel_coupling_matrix}
% 	\end{equation}
%
% \end{widetext}

	\textit{Minimal Hamiltonian eigenbasis}-- Building an accurate low-energy description of the system with the localized $\chi$-orbitals requires a sufficient number $M_{\min{}}$ of Hamiltonian eigenstates $\ket{\psi_i}$ to be initially included in the calculation. To estimate $M_{\min{}}$, we first count how many orbitals belong to the  lowest $K$  shells of a single QD.
	The isotropic parabolic QD in $D$ dimensions with potential energy $\frac{1}{2}m^*\omega^2 \vec{r}^2$ has the spectrum $\hbar\omega\sum_{i=1}^{D}(\nu_i +1/2)$, where $\nu_i =0,1,2,... \forall i$. The states lying within or below the  $K^\text{th}$ degenerate shell satisfy the following condition: 
\begin{equation}
	\nu_1 + \ldots + \nu_D \leq K-1.
\end{equation}
	Introducing $\nu_{D+1} = (K-1) - \sum_{i=1}^{D} \nu_i\geq 0$ transforms the problem into counting the number of nonnegative integers $\nu_i\geq 0$ whose sum is fixed: 
	\begin{equation}
		\# \begin{Bmatrix} \left(\nu_1, \ldots, \nu_{d+1}\right)\in \mathbb{Z}^{D+1}_{\geq 0}:
		\\
		 \sum_{i=1}^{D+1}\nu_i = K-1  \end{Bmatrix}  =
		\begin{pmatrix}
			K-1 + D \\ D
		\end{pmatrix},
		\label{eq:count_degeneracies}
	\end{equation}
as the standard combinatorial ``stars and bars'' method gives. In moderately eccentric QDs, eigenenergies within each shell are distinct but closely clustered together, so the identity~\eqref{eq:count_degeneracies} could still be used as an estimate. 

Now, assume the localization procedure from Eqs.~\eqref{eq:lagrangian}-\eqref{eq:eigenvalue_problem}  has been performed \textit{only} for the $\alpha^\text{th}$ dot  using $M$ lowest-lying Hamiltonian eigenstates $\ket{\psi_i}$. Namely,  the set of $M$ eigenstates $\ket{\xi^{(\alpha)}_i}$ with eigenvalues $\lambda_i^{(\alpha)}$ has been obtained for the overlap matrix $\mathcal{O}^{(\alpha)}$. Using the eigenvalue property of the trace, we can write
\begin{equation}
	\tr(\mathcal{O}^{(\alpha)}) = \sum_{i=1}^{M} \bra{\psi_i} \mathbbm{1}_{{Q}^{(\alpha)}} \ket{\psi_i} = \sum_{i=1}^{M} \lambda_i^{(\alpha)}.
	\label{eq:trace_invariance}
\end{equation}  
In the simplest idealized case, a certain number of $\xi-$states are fully localized within $Q^{(\alpha)}$ with $\lambda_i^{(\alpha)}\approx 1$, and the rest of the states are localized elsewhere with $\lambda_i^{(\alpha)}\approx 0$. Thus, $\tr(\mathcal{O}^{(\alpha)}) $ approximately counts the number of $\xi-$states localized within $Q^{(\alpha)}$. 
Therefore, the number $M=M_{\min{}}$ sufficient to describe at least $K$ shells per QD approximately satisfies the condition: 
\begin{equation}\forall \alpha=1 \ldots  N: \ 
 \sum_{i=1}^{M_{\min{}}} \bra{\psi_i} \mathbbm{1}_{{Q}^{(\alpha)}} \ket{\psi_i} \gtrsim 
		\begin{pmatrix}
			 K-1 + D \\  D
		\end{pmatrix}.
	\label{eq:minimum_subbasis}
\end{equation}
This enables one to estimate $M$  \textit{a priori}--- before performing any localization procedure---by finding the expectation values $\bra{\psi_i} \mathbbm{1}_{{Q}^{(\alpha)}} \ket{\psi_i} $ sequentially for each QD and terminating at $i = M_{\min{}}$ when the condition \eqref{eq:minimum_subbasis} is satisfied.
The exact value of $M$ is problem-specific and is typically above the estimate from \eqref{eq:minimum_subbasis} due to the incompletely spanned subspace of high-energy shells, and is also higher for higher $\lambda_{thr}$ values.

\textit{Disorder simulation}---
For $n$ point defects with charges $q_i=s_i e$ ($s_i=\pm1$) at positions
$(x_i,y_i,z_i)$, the coordinates $x_i$ and $y_i$ are generated from a
uniform random distribution, while $z_i>0$ is generated from an exponential
distribution with mean $2$ nm. The first-order-screened potential energy of these defect charges reads as follows:
\begin{align}
U(x,y)
&=
\sum_{i=1}^{n}U_i^{\mathrm{screened}},
\label{eq:total-screened-energy}
\\
U_i^{\mathrm{screened}}
&=
-s_i\frac{e^2}{4\pi\epsilon_0}A_i
\left(
\frac{1}{R_i}-\frac{1}{R_i'}
\right),
\label{eq:screened-defect-energy}
\\
R_i
&=
\sqrt{
(x-x_i)^2+(y-y_i)^2+(z_e-z_i)^2
},
\\
R_i'
&=
\sqrt{
(x-x_i)^2+(y-y_i)^2+
\left(z_e-\widetilde z_i'\right)^2
}.
\end{align}
Here, $z_e =-1.7 $ nm indicates the plane of charge accumulation in QDs (cf. Fig.\ref{fig:gate_geometry_basis_potential}), $(x_i, y_i, z_i)$ are the positions of defect charges, and $(x_i, y_i, \tilde{z}'_i)$ are the positions of the corresponding image charges, with $\tilde{z}'_i$ found as follows. 
Following \cite{Wilk2001Highkgate}, we approximate the layered oxide by an equivalent SiO$_2$ thickness, defining:
\begin{equation}
h_{\mathrm{eff}}
=
h_{\mathrm{SiO_2}}+\gamma h_{\mathrm{HfO_2}},
\quad
\gamma
=
\frac{\epsilon_{\mathrm{SiO_2}}}
     {\epsilon_{\mathrm{HfO_2}}},
\quad
\widetilde z_i'
=
2h_{\mathrm{eff}}-\widetilde z_i,
\label{eq:effective-image-coordinate}
\end{equation}
where,  $\epsilon_{\mathrm{SiO_2}} = 3.9$, $\epsilon_{\mathrm{HfO_2}} = 25$, and
{\small
\begin{equation}
\widetilde z_i
=
\begin{cases} 
z_i,
&  0\leq z_i<h_{\mathrm{SiO_2}},
\\[4pt] 
h_{\mathrm{SiO_2}}
+\gamma\left(z_i-h_{\mathrm{SiO_2}}\right),
& 
h_{\mathrm{SiO_2}}
\leq z_i
<h_{\mathrm{SiO_2}}+h_{\mathrm{HfO_2}}.
\end{cases}
\label{eq:effective-defect-coordinate}
\end{equation}
}
For the device from Fig.~\ref{fig:gate_geometry_basis_potential}, we use $h_{\mathrm{SiO_2}}=h_{\mathrm{HfO_2}} =10$ nm. The dielectric transmission prefactor $A_i$ depends on the physical layer containing the defect:
{\small
\begin{equation}
A_i
=
\begin{cases}
\displaystyle
\frac{2}
{\epsilon_{\mathrm{SiO_2}}+\epsilon_{\mathrm{Si}}},
&
0\leq z_i<h_{\mathrm{SiO_2}},
\\[10pt]
\displaystyle
\frac{4\epsilon_{\mathrm{SiO_2}} \left(\epsilon_{\mathrm{SiO_2}}+\epsilon_{\mathrm{Si}}\right)^{-1}}
{
\left(\epsilon_{\mathrm{HfO_2}}+\epsilon_{\mathrm{SiO_2}}\right)
},
&
h_{\mathrm{SiO_2}}
\leq z_i
<h_{\mathrm{SiO_2}}+h_{\mathrm{HfO_2}},
\end{cases}
\label{eq:dielectric-transmission-prefactor}
\end{equation}
}
The above follows from the planar-interface coefficient $T_{a \to b} = \frac{2 \epsilon_a}{\epsilon_a + \epsilon_b}$. 

\end{document}